\documentclass[preprint,12pt]{elsarticle}

\usepackage{amssymb}
\usepackage{bm}

\journal{Chemical Physics}

\begin{document}

\begin{frontmatter}

%% Title, authors and addresses

%% use the tnoteref command within \title for footnotes;
%% use the tnotetext command for the associated footnote;
%% use the fnref command within \author or \address for footnotes;
%% use the fntext command for the associated footnote;
%% use the corref command within \author for corresponding author footnotes;
%% use the cortext command for the associated footnote;
%% use the ead command for the email address,
%% and the form \ead[url] for the home page:
%%
%% \title{Title\tnoteref{label1}}
%% \tnotetext[label1]{}
%% \author{Name\corref{cor1}\fnref{label2}}
%% \ead{email address}
%% \ead[url]{home page}
%% \fntext[label2]{}
%% \cortext[cor1]{}
%% \address{Address\fnref{label3}}
%% \fntext[label3]{}

\title{Spreading of waves in nonlinear disordered media}

%% use optional labels to link authors explicitly to addresses:
%% \author[label1,label2]{<author name>}
%% \address[label1]{<address>}
%% \address[label2]{<address>}

\author{S. Flach}

\address{Max-Planck-Institut f\"ur Physik komplexer Systeme,
N\"othnitzer Str. 38, 01187 Dresden, Germany}

\begin{abstract}
%% Text of abstract
We analyze mechanisms and regimes of wave packet spreading in nonlinear disordered media.
We predict that wave packets can spread in two regimes of strong and weak chaos. 
We discuss resonance probabilities, nonlinear diffusion equations, and predict a dynamical
crossover from strong to weak chaos. The crossover is controlled by the ratio of 
nonlinear frequency shifts and the average eigenvalue spacing of eigenstates of the linear equations
within one localization volume. We consider generalized models in higher lattice dimensions
and obtain critical values for the nonlinearity power, the dimension, and norm density, which
influence possible dynamical outcomes in a qualitative way.
\end{abstract}

\begin{keyword}
%% keywords here, in the form: keyword \sep keyword
disorder \sep Anderson localization \sep nonlinearity \sep chaos \sep diffusion

\PACS 05.45-a \sep 05.60Cd \sep 63.20Pw
%% PACS codes here, in the form: \PACS code \sep code

%% MSC codes here, in the form: \MSC code \sep code
%% or \MSC[2008] code \sep code (2000 is the default)

\end{keyword}

\end{frontmatter}

%%
%% Start line numbering here if you want
%%
% \linenumbers

%% main text
\section{Introduction}
\label{Introduction}

In this paper we will discuss the mechanisms of wave packet spreading in nonlinear disordered systems.
More specifically, we will consider cases when i) the corresponding linear wave equations 
yield Anderson localization, ii) the localization length is bounded from above by a finite value,
iii) the nonlinearity is compact in real space and therefore does not induce long range interactions
between eigenstates of the linear equations. 

There are several reasons to analyze such situations. First, wave propagation in spatially disordered
media has been of practical interest since the early times of studies of waves. In particular,
it became of much practical interest for the conductance properties of electrons in semiconductor devices more
than half a century ago. It was probably these issues which motivated P. W. Anderson to perform his
groundbreaking studies on what is now called Anderson localization \cite{PWA58}. With evolving technology, wave propagation
became of importance also in photonic and acoustic devices \cite{Exp,Exp2}. Finally, recent advances in the control over
ultracold atoms in optical potentials made it possible to observe Anderson localization there as well \cite{BECEXP}.
Peter H\"anggi and collaborators studied properties of wave propagation in disordered media
by phase space visualization of the underlying dynamical system in high dimensional phase spaces \cite{hanggi}.

Second, in many if not all cases wave-wave interactions are of importance, or can even be controlled experimentally.
Screening effects can reduce the long range character of these interactions considerably for electrons.
Short range interactions also hold for s-wave scattering of atoms. When many quantum particles interact,
mean field approximations often lead to effective nonlinear wave equations. As a result, nonlinear wave equations
in disordered media become of practical importance. 

Third, there is fundamental interest in understanding, how Anderson localization is modified 
for nonlinear wave equations. All of the above motivates the choice of corresponding
linear wave equations with finite upper bounds on the localization length. Then, the linear equations
admit no transport. Analyzing transport properties of nonlinear disordered wave equations allows to 
observe and characterize the influence of wave-wave interactions on Anderson localization in a straightforward
way.

A number of studies was recently devoted to the above subject 
\cite{Mol98,PS08,Shep08,kkfa08,fks08,skkf09,hvyksf09,mm09,mmkaapds09}. In the present work we will
present a detailed analysis of the chaotic dynamics
which is at the heart of the observed destruction of Anderson localization.
In particular, we will show that an optional intermediate strong chaos regime of subdiffusive
spreading is followed by an even slower subdiffusive spreading process in the regime of weak chaos.

\section{Wave equations}

We will use the Hamiltonian of the disordered discrete nonlinear Schr\"odinger equation
(DNLS)
\begin{equation}
\mathcal{H}_{D}= \sum_{l} \epsilon_{l} 
|\psi_{l}|^2+\frac{\beta}{2} |\psi_{l}|^{4}
- (\psi_{l+1}\psi_l^*  +\psi_{l+1}^* \psi_l)
\label{RDNLS}
\end{equation}
with complex variables $\psi_{l}$, lattice site indices $l$ and nonlinearity
strength $\beta \geq 0$.   The random on-site energies $\epsilon_{l}$
are chosen uniformly from the interval
$\left[-\frac{W}{2},\frac{W}{2}\right]$, with $W$ denoting the disorder
strength.  The equations of motion are generated by $\dot{\psi}_{l} = \partial
\mathcal{H}_{D}/ \partial (i \psi^{\star}_{l})$:
\begin{equation}
i\dot{\psi_{l}}= \epsilon_{l} \psi_{l}
+\beta |\psi_{l}|^{2}\psi_{l}
-\psi_{l+1} - \psi_{l-1}\;.
\label{RDNLS-EOM}
\end{equation}
Eqs.~(\ref{RDNLS-EOM}) conserve the energy (\ref{RDNLS}) and the norm $S
= \sum_{l}|\psi_l|^2$.  We note that varying the norm of an
initial wave packet is strictly equivalent to varying $\beta$.
Eqs.~(\ref{RDNLS}) and (\ref{RDNLS-EOM}) are derived e.~g.~when
describing two-body interactions in ultracold atomic gases on an optical
lattice within a mean field approximation \cite{oberthaler}, but also when
describing the propagation of light through networks of coupled optical
waveguides in Kerr media \cite{yskgpa03}.

Alternatively we also refer to results for the Hamiltonian of the quartic Klein-Gordon lattice (KG) 
\begin{equation}
\mathcal{H}_{K}= \sum_{l}  \frac{p_{l}^2}{2} +
\frac{\tilde{\epsilon}_{l}}{2} u_{l}^2 + 
\frac{1}{4} u_{l}^{4}+\frac{1}{2W}(u_{l+1}-u_l)^2,
\label{RQKG}
\end{equation}
where $u_l$ and $p_l$ are respectively the generalized coordinates and
momenta, and $\tilde{\epsilon}_{l}$ are
chosen uniformly from the interval $\left[\frac{1}{2},\frac{3}{2}\right]$. 
The equations of motion are $\ddot{u}_{l} = - \partial \mathcal{H}_{K}
/\partial u_{l}$ and yield
\begin{equation}
\ddot{u}_{l} = - \tilde{\epsilon}_{l}u_{l}
-u_{l}^{3} + \frac{1}{W} (u_{l+1}+u_{l-1}-2u_l)\;.
\label{KG-EOM}
\end{equation}
Equations (\ref{KG-EOM}) conserve the energy (\ref{RQKG}). They serve e.g. as
simple models for the dissipationless dynamics of anharmonic optical lattice
vibrations in molecular crystals \cite{aaovchinnikov}.  The energy of an
initial state $E \geq 0$ serves as a control parameter of nonlinearity similar
to $\beta$ for the DNLS case.
For small amplitudes the equations of motion of the KG chain can be
approximately mapped onto a DNLS model
\cite{KG-DNLS-mapping}. For the KG model with given parameters $W$ and $E$, the corresponding
DNLS model (\ref{RDNLS}) with norm $S=1$, has a nonlinearity parameter
$\beta\approx 3WE$. 
The norm density of the DNLS model corresponds to the normalized energy
density of the KG model.

The theoretical considerations will be performed within the DNLS framework.
It is straightforward to adapt them to the KG case.

\section{Anderson localization}

For $\beta=0$ with $\psi_{l} = A_{l}
\exp(-i\lambda t)$ Eq.~(\ref{RDNLS})
is reduced to the linear eigenvalue problem
\begin{equation}
\lambda A_{l} = \epsilon_{l} A_{l} 
- A_{l-1}-A_{l+1}\;.
\label{EVequation}
\end{equation}
The normalized eigenvectors $A_{\nu,l}$ ($\sum_l A_{\nu,l}^2=1)$ are the NMs,
and the eigenvalues $\lambda_{\nu}$ are the frequencies of the NMs.  The width
of the eigenfrequency spectrum $\lambda_{\nu}$ of (\ref{EVequation}) is
$\Delta=W+4$ with $\lambda_{\nu} \in \left[ -2 -\frac{W}{2}, 2 + \frac{W}{2}
\right] $.

The asymptotic spatial decay of an eigenvector is given by $A_{\nu,l} \sim
{\rm e}^{-l/\xi(\lambda_{\nu})}$ where 
$\xi(\lambda_{\nu})$ is the localization length and
$\xi(\lambda_{\nu}) \approx
24(4-\lambda_{\nu}^2)/W^2$ for weak disorder $W \leq 4$ \cite{PWA58,KRAMER}. The NM participation number
$p_{\nu} = 1/\sum_l A_{\nu,l}^4$ is one possible way to quantize the spatial extend
(localization volume) of the NM. The localization volume $V$ is on average of the order of $3 \xi(0)$
for weak disorder, and tends to $V=1$ in the limit of strong disorder.
The average spacing $d$ of eigenvalues of NMs
within the range of a localization volume is therefore of the order of $d \approx \Delta / V$,
which becomes $d \approx \Delta W^2 /300 $ for weak disorder.
The two scales $ d \leq \Delta $ are expected to determine the
packet evolution details in the presence of nonlinearity.

Due to the localized character of the NMs, any localized wave packet with size $L$ which is launched into
the system for $\beta=0$ , will stay localized for all times. If $L \ll V$, then the wave packet will expand
into the localization volume. This expansion will take a time of the order of $\tau_{lin}=2\pi/d$. 
If instead $L \geq V$, no substantial expansion will be observed in real space.
We remind that Anderson localization is relying on the phase coherence of waves. Wave packets which are trapped due
to Anderson localization correspond to trajectories in phase space evolving on tori, i.e. quasiperiodically in time.

\section{Adding nonlinearity}

The equations of motion of (\ref{RDNLS-EOM})  in normal mode space read
\begin{equation}
i \dot{\phi}_{\nu} = \lambda_{\nu} \phi_{\nu} + \beta \sum_{\nu_1,\nu_2,\nu_3}
I_{\nu,\nu_1,\nu_2,\nu_3} \phi^*_{\nu_1} \phi_{\nu_2} \phi_{\nu_3}\;
\label{NMeq}
\end{equation}
with the overlap integral 
\begin{equation}
I_{\nu,\nu_1,\nu_2,\nu_3} = 
\sum_{l} A_{\nu,l} A_{\nu_1,l} 
A_{\nu_2,l} A_{\nu_3,l}\;.
\label{OVERLAP}
\end{equation}
The variables $\phi_{\nu}$ determine the complex time-dependent amplitudes of
the NMs.

The frequency shift of a single site oscillator induced by the nonlinearity is
$\delta_l = \beta |\psi_l|^{2}$. If instead a single mode is excited, its
frequency shift can be estimated by $\delta_{\nu} = \beta |\phi_{\nu}|^2/
p_{\nu}$. 

As it follows from (\ref{NMeq}), nonlinearity induces an interaction between NMs.
Since all NMs are exponentially localized in space, each normal mode is effectively coupled to a finite
number of neighbouring NMs, i.e. the interaction range is finite. However the strength of the coupling
is proportional to the norm density $n = |\phi|^2$. Let us assume that a wave packet spreads.
In the course of spreading its norm density will become smaller. Therefore the effective coupling strength
between NMs decreases as well. At the same time the number of excited NMs grows.

One possible outcome would be: (I) that after some time the coupling will be weak enough
to be neglected. If neglected, the nonlinear terms are removed, the problem is reduced to the linear wave equation,
and we obtain again Anderson localization. That implies that the trajectory happens to be on a quasiperiodic torus.
Then it has to be on that torus from the beginning. 
Another possibility is: (II) that spreading continues for all times. That would imply that the trajectory
evolves not on a quasiperiodic torus, but in some chaotic part of phase space.
A third possibility is: (III) that the trajectory was initially strongly chaotic, but manages in the course of spreading
to get trapped between denser and denser torus structures in phase space after some spreading, leading again
to localization as an asymptotic outcome.

Consider a wave packet with size $L$ and norm density $n$. Replace it by a {\sl finite} system of size $L$ and
norm density $n$. 
Such a finite system will be in general nonintegrable. Therefore the only possibility
to generically obtain a quasiperiodic evolution is to be in the regime where the KAM theorem holds.
Then there is a finite fraction of the available phase space volume which is filled with KAM tori.
For a given $L$ it is expected that there is a critical density $n_{KAM}(L)$ below which the KAM regime will hold.
We do not know this $L$-dependence. Computational studies may not be very conclusive here, since it is hard to distinguish
a regime of very weak chaos from a strict quasiperiodic one on finite time scales.

The above first possible outcome (I) (localization) will be realized if the packet is launched in a KAM regime.
Whether that is possible at all for an infinite system is an open issue.
The second outcome (II) (spreading) implies that we start in a chaotic regime and remain there. Since 
the packet density is reduced and is proportional to its inverse size $L$ at later times, 
this option implies that the critical density $n_{KAM}(L)$ decays faster than $1/L$, possibly faster than any
power of $1/L$.
The third possibility (III) (asymptotic localization) should be observable by some substantial slowing down of the
spreading process.

\subsection{The secular normal form}

Let us perform a further transformation $\phi_{\nu} = {\rm e}^{-i \lambda_{\nu} t} \chi_{\nu}$ and insert
it into Eq. (\ref{NMeq}):
\begin{equation}
i \dot{\chi}_{\nu} = \beta \sum_{\nu_1,\nu_2,\nu_3}
I_{\nu,\nu_1,\nu_2,\nu_3} \chi^*_{\nu_1} \chi_{\nu_2} \chi_{\nu_3} {\rm e}^{i(\lambda_{nu} + 
\lambda_{\nu_1}-\lambda_{\nu_2}-\lambda_{nu_3})t}
\;.
\label{NMeqchi}
\end{equation}
The right hand side contains oscillating functions with frequencies
\begin{equation}
\lambda_{\nu,\vec{\nu}} \equiv \lambda_{\nu} + 
\lambda_{\nu_1}-\lambda_{\nu_2}-\lambda_{\nu_3}\;,\;\vec{\nu} \equiv (\nu_1,\nu_2,\nu_3)\;.
\label{dlambda}
\end{equation}
For certain values of $\nu,\vec{\nu}$ the value $\lambda_{\nu,\vec{\nu}}$ becomes exactly zero. These secular terms
define some slow evolution of (\ref{NMeqchi}). Let us perform an averaging over time of all terms
in the rhs of (\ref{NMeqchi}), leaving therefore only the secular terms. The resulting secular
normal form equations (SNFE) take the form
\begin{equation}
i \dot{\chi}_{\nu} = \beta \sum_{\nu_1}
I_{\nu,\nu_,\nu_1,\nu_1} |\chi_{\nu_1}|^2 \chi_{\nu}
\;.
\label{NMeqRNF}
\end{equation}
Note that possible missing factors due to index permutations can be absorbed into the overlap integrals,
and are not of importance for what is following.
The SNFE can be now solved for any initial condition $\chi_{\nu}(t=0)=\eta_{\nu}$ and yield
\begin{equation}
\chi_{\nu}(t) = \eta_{\nu} {\rm e}^{-i \Omega_{\nu} t}\;,\; \Omega_{\nu} = \beta \sum_{\nu_1}
I_{\nu,\nu_,\nu_1,\nu_1} |\eta_{\nu_1}|^2 
\;.
\end{equation}
Since the norm of every NM is preserved in time for the SNFE, it follows that Anderson localization
is preserved within the SNFE. The only change one obtains is the renormalization of the eigenfrequencies
$\lambda_\nu$ into $\tilde{\lambda}_{\nu} = \lambda_{\nu}+\Omega_{\nu}$. Moreover, the phase coherence
of NMs is preserved as well. Any different outcome will be therefore due to the nonsecular terms,
neglected within the SNFE.

\subsection{Measuring properties of wave packets}

We order the NMs in space by increasing value of the center-of-norm coordinate
$X_{\nu}=\sum_l l A_{\nu,l}^2$.  We analyze normalized distributions $n_{\nu}
\geq 0$ using the second moment $m_2= \sum_{\nu}
(\nu-\bar{\nu})^2 n_{\nu}$, which quantifies the wave packet's degree of
spreading and the participation number $P=1 / \sum_{\nu} n_{\nu}^2$, which
measures the number of the strongest excited sites in $n_{\nu}$.  Here
$\bar{{\nu}} = \sum_{\nu} \nu n_{\nu}$.  We follow norm density
distributions $n_{\nu}\equiv |\phi_{\nu}|^2/\sum_{\mu} |\phi_{\mu}|^2$.
The second moment $m_2$ is sensitive to the distance of the tails of
a distribution from the center, while the participation number $P$ is a measure of the
inhomogeneity of the distribution, being insensitive to any spatial
correlations. Thus, $P$ and $m_2$ can be used to quantify the
sparseness of a wave packet through the compactness index 
\begin{equation}
\zeta=\frac{P^2}{m_2}.
\label{eq:ci}
\end{equation} 
A thermalized wave packet yields $\zeta=3$.
Distributions with larger gaps between 
equally excited isolated sites attain a compactness index $\zeta<3$. 

\subsection{Expected regimes of spreading}

Previous studies suggested the existence of various dynamical regimes of spreading of wave packets
\cite{PS08,fks08,skkf09}. Some of these
definitions were contradictory. Below we will resolve this.

Consider a wave packet at $t=0$ which has norm density $n$ and size $L$.
If $\beta n \geq \Delta$, a substantial part of the wave packet will be selftrapped \cite{kkfa08,skkf09}.
This is due to the above discussed nonlinear frequency shifts, which will tune the excited sites
immediately out of resonance with the nonexcited neighborhood. As a result, discrete breather like structures will
be formed, which can persist for immensely long times. While selftrapping and discrete breather formation
are interesting localization phenomena at strong nonlinearity, they are very different from Anderson localization
since they require the existence of gaps in the spectrum of the linear wave equations \cite{DB}.
If now $\beta n < \Delta$, selftrapping is avoided, and the wave packet can start to spread.
For $L < V$ and $\beta=0$, the packet will spread over the localization volume during the time $\tau_{lin}$. 
After that, the new norm density will drop down to $n(\tau_{lin}) \approx n \frac{L}{V}$. For $L > V$ the norm density will not
change appreciably up to $\tau_{lin}$, $n(\tau_{lin}) \approx n$. 
The nonlinear frequency shift $\beta n(\tau_{lin})$ can be now compared with the average spacing $d$.
If $\beta n(\tau_{lin}) > d$, all NMs in the packet are resonantly interacting with each other. This regime
will be coined strong chaos. If instead $\beta n(\tau_{lin}) < d$, NMs are weakly interacting with each other.
This regime will be coined weak chaos. To summarize:
\begin{eqnarray}
\beta n(\tau_{lin}) < d \;:\; {\rm weak\; chaos}
\nonumber
\\
\beta n(\tau_{lin}) > d \;:\; {\rm strong\; chaos}
\nonumber
\\
\beta n > \Delta \;:\; {\rm selftrapping}
\nonumber
\end{eqnarray}
In terms of the above wave packet characteristics $n$ , $L$ it follows
\begin{eqnarray}
\beta n \tilde{L} < \Delta \;:\; {\rm weak\; chaos}
\nonumber
\\
\beta n \tilde{L} > \Delta \;:\; {\rm strong\; chaos}
\label{regimes}
\\
\beta n > \Delta \;:\; {\rm selftrapping}
\nonumber
\end{eqnarray}
where $\tilde{L}=L$ for $L < V$ and $\tilde{L}=V$ for $L > V$. 
It follows that the regime of strong chaos can be observed only if $L > 1$. For $L=1$
we expect only two regimes - selftrapping and weak chaos.
Furthermore, we obtain that the regimes of strong and weak chaos are separated by the quantity
$\beta n = d$, i.e. the average spacing $d$ is the only characteristic frequency scale here.

\subsection{Discussion of numerical results}

\begin{figure}
\includegraphics[angle=0,width=0.99\columnwidth]{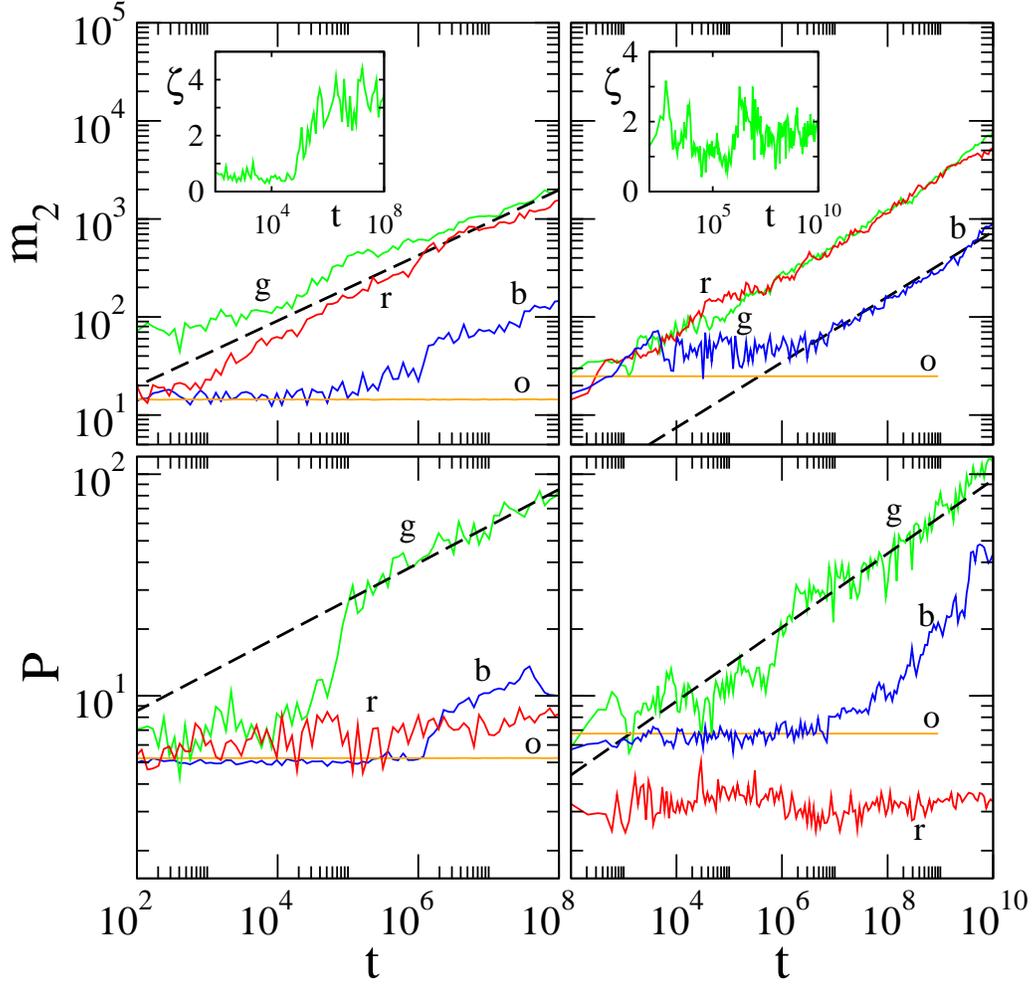}
\caption{(color online) Single site excitations.  $m_2$ and $P$ versus time in
log--log plots.  Left plots: DNLS with $W=4$, $\beta=0,0.1,1,4.5$ [(o),
orange; (b), blue; (g) green; (r) red].  Right plots: KG with $W=4$ and
initial energy $E=0.05,0.4,1.5$ [(b) blue; (g) green; (r) red]. The orange
curves (o) correspond to the solution of the linear equations of motion, where
the term $u_l^3$ in (\ref{KG-EOM}) was absent.  The disorder realization is
kept unchanged for each of the models.  Dashed straight lines guide the eye
for exponents 1/3 ($m_2$) and 1/6 ($P$) respectively. Insets: the compactness
index $\zeta$ as a function of time in linear--log plots for $\beta=1$ (DNLS)
and $E=0.4$ (KG). Adapted from \cite{skkf09}}.
\label{fig1}
\end{figure}

Let us discuss the above in the light of published computational experiments.
We show results for single site excitations from \cite{skkf09} in Fig.\ref{fig1} with $W=4$, $L=1$ and $n=1$.
%&&&&&&&&&&&&&&&&&&&&&&&&&&&&&&&&&&&&&&&&&
%&&&&&&&&&&&&&&&&&&&&&&&&&&&&&&&&&&&&&&&&&
For the DNLS model (left plots in Fig.\ref{fig1}) with $\beta=4.5$ it follows
$\beta n = 4.5$. Already at these values selftrapping of a part of the wave packet is observed.
Therefore $P$ does not grow significantly, while the second moment $m_2\sim
t^{\alpha}$ with $\alpha \approx 1/3$ (red curves).  A part of the
excitation stays highly localized \cite{kkfa08}, while another part
delocalizes. 
For $\beta=1$ selftrapping is avoided since  $\beta n < \Delta$. With $V\approx 20$ and
$d \approx 0.4$ it follows that $\tau_{lin} \approx 16$ and $\beta n(\tau_{lin}) \approx 0.05 \ll d$.
Therefore we expect to observe the regime of weak chaos. It is characterized
by subdiffusive spreading with $m_2\sim t^{\alpha}$
and $P \sim t^{\alpha/2}$ (green curves).  
For $\beta=0.1$ we will remain in the regime of weak chaos, however the time scales for observing spreading
grow. Therefore one finds
no visible spreading up to some time  $\tau_d$ which increases with further
decreasing nonlinearity. For $t < \tau_d$ both $m_2$ and $P$ are not changing.
However for $t > \tau_d$ the packet shows visible growth
with the characteristics of weak chaos (blue curves).  The
simulation of the equations of motion in the absence of nonlinear terms (orange
curves) shows Anderson localization.
Since $L=1$ in the above numerical data, strong chaos has not been observed.

Notably, the authors of Ref. \cite{skkf09} also considered single mode excitations with total norm $S=1$.
Using the above terminology, $n\approx 1/V$ and $L=V$ with $W=4$ and therefore again $V\approx 20$.
For the case $\beta=6.5$ the authors detected a growth of $m_2$ which was subdiffusive but faster than
$t^{1/3}$. We think that these observations are a clear hint towards the realization of strong chaos,
which should be observable for $5...10 < \beta < 30...40$ in these cases.

The time evolution of
$\zeta$ for excitations in the regime of weak chaos is shown in the insets of
Fig.~\ref{fig1}. As one can see the compactness index oscillates around some
constant nonzero value both for the DNLS and the KG models. This means that
the wave packet spreads but does not become more
sparse. 
The average
value $\overline{\zeta}$ of the compactness index over 20 realizations
of single mode excitations
at $t=10^8$ for the DNLS model with $W=4$ and $\beta=5$ was found to be
$\overline{\zeta}=2.95 \pm 0.39$ \cite{skkf09}.
%&&&&&&&&&&&&&&&&&&&&&&&&&&&&&&&&&&&&&&&&&
\begin{figure}
\includegraphics[angle=0,width=0.99\columnwidth]{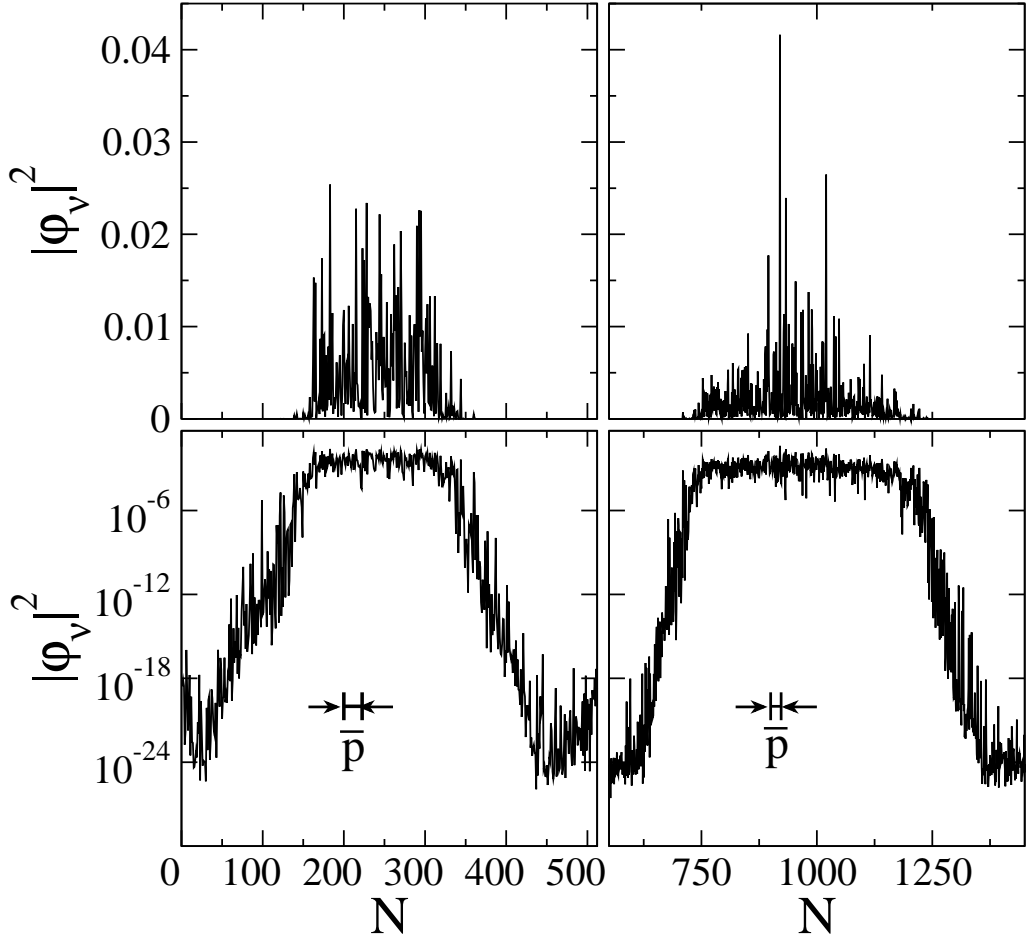}
\caption{Norm density distributions in the NM space at time
$t=10^8$ for the initial excitations in the regime of weak chaos of the DNLS model. 
Left plots: single
site excitation for $W=4$ and $\beta=1$. Right plots: single mode
excitation for $W=4$ and $\beta=5$.  $|\phi_\nu|^2$ is plotted in
linear (logarithmic) scale in the upper (lower) plots. The average
localization volume $V\approx20$ (shown
schematically in the lower plots) is much smaller than the length over
which the wave packets have spread. Adapted from \cite{skkf09}.  }
\label{fig2}
\end{figure}
%&&&&&&&&&&&&&&&&&&&&&&&&&&&&&&&&&&&&&&&&&

The norm density distribution for the
DNLS model at $t=10^8$ is plotted in Fig.~\ref{fig2}.  The distribution is characterized by a flat plateau of
almost ideally thermalized NMs. The width of this plateau is more than an order of magnitude
larger than the localization volume of the linear equations. Therefore Anderson localization
is destroyed. The plateau is bounded by exponentially decreasing tails, with exponents corresponding
to the localization length of the linear equations. With growing time the plateau widens, drops in height,
and is pushing the tails to larger distances.
Another remarkable feature are the huge fluctuations of norm densities in the tails, reaching 4-6 orders
of magnitude. Such fluctuations are observed even in the case $\beta=0$. They are due to the fact,
that NMs are ordered in space. Neighbouring NMs in space may have different eigenfrequencies, and therefore
different values of their localization length. Tail NMs are excited by the core. The further away they are,
the weaker the excitation. But within a small tail volume, NMs with larger localization length will be 
more strongly excited than those with smaller localization length, hence the large observed fluctuations,
which on a logarithmic scale are of the order of the relative variation of the localization length.
The remarkable observation is, that these fluctuations in the tails persist for the nonlinear case.
Anderson localization is destroyed in the core (plateau) of the wave packet due to mode-mode interactions.
The tail NMs are slaved to the core and excited by it. The interaction between neighbouring tail NMs
is negligible, and the huge fluctuations persist. Therefore, Anderson localization is preserved in the tails
of the distributions over very long times (essentially until the given tail volume becomes a part of
the core).

For single site
excitations in the regime of weak chaos the exponent $\alpha$ of subdiffusive spreading does not appear to depend on $\beta$ in the
case of the DNLS model or on the value of $E$ in the case of KG. We find no visible dependence of the
exponent $\alpha$ on $W$.  Therefore the subdiffusive spreading is rather
universal and the parameters $\beta$ (or $E$) and $W$ are only affecting the
prefactor.  Excluding selftrapping, any nonzero
nonlinearity  appears to completely delocalize the wave packet and destroy Anderson
localization.  Fittings were performed by analyzing 20 runs in the regime of weak chaos with
different disorder realizations. For each realization the exponent
$\alpha$ was fitted, and then averaged over all computational measurements.  We find
$\alpha = 0.33 \pm 0.02$ for DNLS, and $\alpha = 0.33 \pm 0.05$ for KG \cite{fks08,skkf09}.
Therefore, the universal exponent $\alpha=1/3$ \cite{fks08} appears
to explain the data.

Another intriguing test was performed on the same disorder realizations and single site initial conditions,
by additionally dephasing the NMs in a random way every hundred time units \cite{skkf09}. In that case, subdiffusion
speeds up, and $m_2$ grows as $t^{1/2}$ implying $\alpha_{deph} = 1/2$. This regime of complete decoherence of
NM phases exactly corresponds to the above anticipated one of strong chaos, but here enforced by explicit dephasing.

\section{From strong to weak chaos, from resonances to nonlinear diffusion}

We can think of two possible mechanisms of wave packet spreading. A NM with
index $\mu$ in a boundary layer of width $V$ in the cold exterior, which
borders the packet, is either incoherently {\sl heated} by the packet, or {\sl
resonantly excited} by some particular NM from a boundary layer with width
$V$ inside the packet. 

For heating to work, the packet modes $\phi_{\nu}(t)$ should contain a part
$\phi_{\nu}^{c}(t)$, having a continuous frequency spectrum (similar to a
white noise), in addition to a regular part $\phi_{\nu}^{r}(t)$ of pure point
frequency spectrum:
\begin{equation}
\phi_{\nu}(t) = \phi_{\nu}^{r}(t) + \phi_{\nu}^{c}(t)\;.
\label{cpp}
\end{equation}
Therefore at least some NMs of the packet should evolve chaotically in time.
The more the packet spreads, the less the mode amplitudes in the packet
become. Therefore its dynamics should become more and more regular, implying $
\lim_{t\rightarrow \infty} \phi_{\nu}^{c}(t)/\phi_{\nu}^{r}(t) \rightarrow 0$.

%_______________________________________________
\subsection{Strong chaos}
\label{sec:dephase}

Let us assume that all NMs in the packet are strongly chaotic, and
their phases can be assumed to be random on the time scales of the observed spreading.  
According to (\ref{NMeq}) the heating of the exterior mode should evolve as $i
\dot{\phi}_{\mu} \approx \lambda_{\mu} \phi_{\mu} + \beta n^{3/2} f(t)$ where
$\langle f(t) f(t') \rangle = \delta(t-t')$ ensures that $f(t)$ has a
continuous frequency spectrum.  Then the exterior NM  increases its norm
according to $|\phi_{\mu}|^2 \sim \beta^2 n^3 t$. The momentary diffusion rate
of the packet is given by the inverse time $T$ it needs to heat the exterior
mode up to the packet level: $D = 1/T \sim \beta^2 n^2$. 
The second moment is of the order of $m_2 \sim 1/n^2$, since the packet size is $1/n$. The diffusion
equation $m_2 \sim D t$ yields $m_2 \sim \beta t^{1/2}$.  
This agrees very well with the numerical results for dephasing in NM space.
Moreover, we expect it to hold also without explicit dephasing, provided the initial wave packet
satisfies the above conditions for strong chaos (\ref{regimes}). 
First numerical tests show that this is correct \cite{tljb10},
but it contradicts the observations of the numerical data in the regime of weak chaos without additional dephasing.
Thus, in the regime of weak chaos, not all
NMs in the packet are chaotic, and dephasing is at best some partial outcome.

\subsection{Resonance probability}

Chaos is a combined result of resonances and nonintegrability. Let us estimate
the number of resonant modes in the packet for the DNLS model.  Excluding
secular interactions, the amplitude of a NM with
$|\phi_{\nu}|^2 = n_{\nu}$ is modified by a triplet of other modes 
$\vec{\mu}\equiv (\mu_1,\mu_2,\mu_3)$ in first order in $\beta$ as (\ref{NMeq})
\begin{equation}
|\phi_{\nu}^{(1)}| = \beta \sqrt{n_{\mu_1}n_{\mu_2}n_{\mu_3}}
R_{\nu,\vec{\mu}}^{-1}\;,\; R_{\nu,\vec{\mu}} \sim
\left|\frac{\lambda_{\nu,\vec{\mu}}}{I_{\nu,\mu_1,\mu_2,\mu_3}}\right| \;,
\label{PERT1}
\end{equation}
where $\lambda_{\nu,\vec{\mu}} =
\lambda_{\nu}+\lambda_{\mu_1}-\lambda_{\mu_2}-\lambda_{\mu_3}$.  The
perturbation approach breaks down, and resonances set in, when $\sqrt{n_{\nu}}
< |\phi_{\nu}^{(1)}|$.  Since all considered NMs belong to the packet, we
assume their norms to be equal to $n$ for what follows.  
Then the resonance condition for a given NM with index $\nu$ 
is met if there is at least one given triplet of other NMs $\vec{\mu}$ such that
\begin{equation}
\beta n < R_{\nu,\vec{\mu}}\;.
\label{resonance_R}
\end{equation}
If three of the four
mode indices are identical, one is left with interacting NM pairs.  A
statistical analysis of the probability of resonant interaction was performed
in Ref. \cite{fks08}.  For small values of $n$ (i.e. when the packet has
spread over many NMs) the main contribution to resonances are due to rare
multipeak modes \cite{fks08}, with peak distances being larger than the
localization volume. However pair resonances are expected not to contribute to
the spreading process \cite{hvyksm10}. When distances between the peaks of multipeak modes are larger
than the localization volume, the time scale of excitation transfer from one peak to another
will grow exponentially with the distance. Such processes are too slow in order to
be observed in numerical experiments \cite{hvyksm10}.

If two or none of the four mode indices are identical,
one is left with triplets and quadruplets of interacting NMs respectively. In
both cases the resonance condition (\ref{resonance_R}) can be met at arbitrarily small values of
$n$ for NMs from one localization volume.

For a
given NM $\nu$ we define $ R_{\nu,\vec{\mu}_0} = \min_{\vec{\mu} }
R_{\nu,\vec{\mu}}$.  Collecting $R_{\nu,\vec{\mu}_0}$ for many $\nu$ and many
disorder realizations, we can obtain the probability density distribution
$\mathcal{W}(R_{\nu,\vec{\mu}_0})$. 
The probability $\mathcal{P}$ for a mode, which is excited to a norm $n$ (the
average norm density in the packet), to be resonant with at least one triplet
of other modes at a given value of the interaction parameter $\beta$ is therefore given
by
\begin{equation}
\mathcal{P} = \int_0^{\beta n} \mathcal{W}(x) {\rm d}x\;.
\label{resprob}
\end{equation} 
The main result is that $\mathcal{W}(R_{\nu,\vec{\mu}_0} \rightarrow 0)
\rightarrow C(W) \neq 0$ \cite{skkf09}.  For the cases studied, the constant $C$ drops with
increasing disorder strength $W$. This result of nonzero values of $C$ 
is not contradicting the fact of level repulsion
of neighbouring NMs, since triplet and quadruplet combinations of NM frequencies can yield
practically zero values of $\lambda_{\nu,\vec{\mu}}$ with finite distances between the eigenfrequencies.

For the case of strong disorder ($W \gg 1$) the localization volume tends to one,
and quadruplet resonances are rare. Excluding also pair resonances for the above reasons,
we are left with triplet resonances. A given mode may yield a triplet resonance with
its two nearest neighbours to the left and right. Replacing the overlap integrals by some characteristic
average, and assuming that the three participating modes have essentially uncorrelated eigenfrequencies,
it follows that 
\begin{equation}
\mathcal{W}(R) \approx C \left( 1-\frac{CR}{3} \right) ^2
\;.
\end{equation}
Due to the nonnegativity of $\mathcal{P}$ it would follow that $\mathcal{P}=0$ for $R \geq 3/C$.
In reality we expect an exponential tail for large $R$. As a simple approximation, we may instead
use
\begin{equation}
\mathcal{W} (R) \approx C {\rm e}^{-CR}
\label{approxp}
\end{equation}
which in turn can be expected to hold also for the case of weak disorder.
It leads to the approximative result
\begin{equation}
\mathcal{P} = 1-{\rm e}^{-C\beta n}\;.
\label{approxpp}
\end{equation}

For $\beta n \rightarrow 0$ it follows
\begin{equation}
\mathcal{P} \approx C \beta n\;.
\label{resprobas}
\end{equation}
Therefore the probability for a mode in the packet to be resonant is
proportional to $C \beta n$ in the limit of small $n$ \cite{fks08,skkf09}.  However, on average the number of resonant modes in the
packet is proportional to the product of $\mathcal{P}$ and the total number of modes in the packet.
Since the total number is proportional to $1/n$, the the average number of resonant
modes in a packet is constant, proportional to $C \beta$, and their fraction within the
packet is $\sim C \beta n$ \cite{fks08,skkf09}.  Since packet mode amplitudes fluctuate in
general, averaging is meant both over the packet, and over suitably long time
windows. A detailed numerical analysis of the statistical properties of resonances and related issues is
in preparation \cite{dksf10}.

Finally we consider the process of resonant excitation of an exterior mode by
a mode from the packet. The number of packet modes in a layer of the width of
the localization volume at the edge, which are resonant with a cold exterior
mode, will be proportional to $\beta n$. After long enough spreading $\beta n
\ll 1$.  On average there will be no mode inside the packet, which could
efficiently resonate with an exterior mode. Resonant growth can be
excluded \cite{fks08,skkf09}.
Thus, a wave packet is trapped at its edges, and stays localized until
the interior of the wave packet decoheres (thermalizes). On these (growing) time scales,
the packet will be finally able to incoherently excite the exterior and to extend its size.

\subsection{A conjecture leading to the correct asymptotics}

We assume, that the continuous frequency part of the dynamics of a packet mode
is $\mathcal{P}(\beta n)$.
It follows that $\phi_{\nu}^{c}(t) / \phi_{\nu}^{r}(t) \sim \mathcal{P}(\beta n)$. 
As expected initially, the chaotic part in the dynamics of packet modes
becomes weaker the more the packet spreads, and the packet dynamics
becomes more and more regular in the limit of large times.  Therefore the
chaotic component is conjectured to be a small parameter $\phi_{\nu}^{c}(t) \ll \phi_{\nu}^{r}(t)$.
Expanding the term $|\phi_{\nu}|^2\phi_{\nu}$ to first order in
$\phi_{\nu}^{c}(t)$, the heating of the exterior mode should evolve according
to $i \dot{\phi}_{\mu} \approx \lambda_{\mu} \phi_{\mu} + \beta n^{3/2} \mathcal{P}(\beta n)
f(t)$.  It follows $|\phi_{\mu}|^2 \sim  \beta^2 n^3 (\mathcal{P}(\beta n))^2 t$, and the rate 
\begin{equation}
D =
1/T \sim  \beta^2 n^2 (\mathcal{P}(\beta n))^2\;.
\end{equation}
%(cf.~the prediction (\ref{diffcoef})). 
With (\ref{approxp}),(\ref{approxpp}) and $m_2 \sim 1/n^2$
the
diffusion equation $m_2 \sim D t$ yields
\begin{equation}
\frac{1}{n^2} \sim \beta (1-{\rm e}^{-C\beta n }) t^{1/2}\;.
\label{strong-weak}
\end{equation}
The solution of this equation yields a crossover from subdiffusive spreading in the regime of
strong chaos to subdiffusive spreading in the regime of weak chaos:
\begin{eqnarray}
m_2 \sim (\beta^2 t)^{1/2}\;,\;{\rm strong}\;{\rm chaos}\;,\;C\beta n > 1\;,
\nonumber
\\
m_2 \sim C^{2/3} \beta^{4/3} t^{1/3}\;,\;{\rm weak}\;{\rm chaos}\;,\;C\beta n < 1\;,
\nonumber
\end{eqnarray}

\subsection{The crossover from strong chaos to weak chaos}

According to (\ref{resprob}) the probability of resonance for a packet NM will be practically equal to one,
if $\beta n$ is sufficiently larger than $1/C$. Such a situation can be generated for packets with large enough 
$\beta n$, and should yield spreading, provided one avoids selftrapping $\beta n \leq 4+W$ \cite{kkfa08,skkf09}. 
This spreading will be different from the asymptotic behaviour discussed above over potentially large time scales.

Let us use as an example $W=4$ and $\beta=1$, with the constant $C\approx 6.2$ \cite{skkf09}. 
Single site excitations with norm $S=1$
lead after very short times to a spreading of the excitation into the localization volume of the linear wave equations,
which is of the order of 10-20.
The attained norm density is therefore of the order of $n \leq 0.1$. 
The observed spreading is the asymptotic one since $\mathcal{P} \sim  C \beta n$. However, if we choose a packet
size $L$ to be of
the order of the localization volume, and the norm density $n$ of the order of $n=1$, initially $\mathcal{P} \approx 1$. 
Thus every mode in the packet will be resonant, and the condition for strong chaos should hold.
At the same time $\beta n=1$ is far below the selftrapping threshold $4+W = 8$.
For strong chaos we derived $m_2 \sim t^{1/2}$. With spreading continuing, the norm density in the packet will
decrease, and eventually $\beta n \leq 1/C$. Then there will be a crossover from strong chaos to weak chaos,
and $m_2\sim t^{1/3}$ for larger times. This crossover happens on logarithmic time scales, and it will be not easy to
confirm it numerically \cite{tljb10}. 

\begin{figure}
\includegraphics[angle=-90,width=0.99\columnwidth]{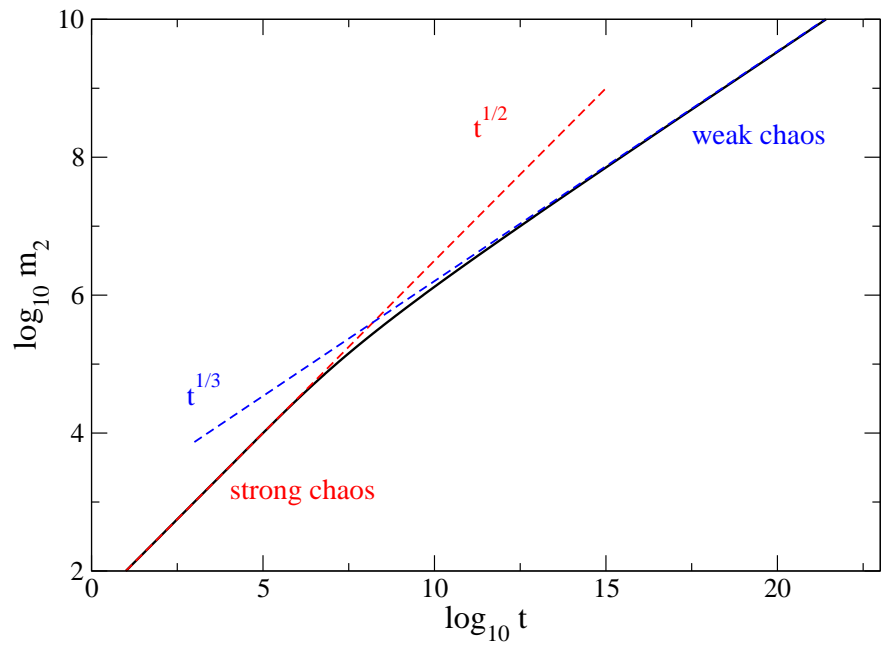}
\caption{$m_2(t)$ in a log-log plot according to (\ref{strong-weak}) (black solid line).
Dashed lines - power laws for strong and weak chaos.}
\label{fig3}
\end{figure}

In Fig.\ref{fig3} we show the resulting time dependence of $m_2$ on $t$ from (\ref{strong-weak}) in a log-log plot,
where we used $\beta=1$, $C=6.2$, $L=20$ and $n(t=10^2)=1$.
%&&&&&&&&&&&&&&&&&&&&&&&&&&&&&&&&&&&&&&&&&
%&&&&&&&&&&&&&&&&&&&&&&&&&&&&&&&&&&&&&&&&&
With $x=log_{10}(t)$ and $y=log_{10}(m_2)$ it is straightforward
to calculate the zero of the third derivative $d^3y/dx^3=0$ to obtain the crossover position:
\begin{equation}
C\beta n_c \approx 1.86\;.
\end{equation}
Therefore the only characteristic frequency scale here is $1/C$. From the above discussion of the different
spreading regimes (\ref{regimes}) it follows, that this scale is corresponding to the average spacing $d$:
\begin{equation}
\frac{1}{C} \approx d\;.
\end{equation}
Then
\begin{eqnarray}
C \approx \frac{100}{W^2}\;,\;W \leq 4 \;,
\\
C \approx \frac{1}{W}\;,\; W \gg 4\;.
\end{eqnarray}
Our results can be used to predict the critical value of the norm density $n_c$ at which the crossover should
take place. For $W=4$ and $\beta=1$ it follows $n_c\approx 0.3$.

\subsection{Scaling of nonlinear diffusion equations}

With the above results on the diffusion coefficient, we may consider a set of nonlinear diffusion
equations for the norm density distributions in NM space. For simplicity we replace the discrete
NM indices by a continuous variable:
\begin{equation}
\frac{\partial n}{\partial t} = \frac{\partial}{\partial \nu} D(n) \frac{\partial n}{\partial \nu} 
\;.
\end{equation}
In particular, we are interested in cases $D(n) \sim n^{\kappa}$. With a single scaling assumption
$n(\nu,t/a) = b n(c\nu,t)$, and using the conservation of the total norm and $n(\nu \rightarrow \pm \infty,t)
\rightarrow 0$ we obtain $b=c=a^{1/(\kappa+2)}$. Then the second moment $m_2$ will grow in time 
according to
\begin{equation}
m_2(t) = \left( \frac{t}{t_0}\right) ^{\alpha} m_2(t_0)\;,\;\alpha=\frac{2}{\kappa+2}\;.
\end{equation}

Notably an explicit self-similar solution was calculated by Tuck in 1976 \cite{tuck76} which has
the following spatial profile:
\begin{equation}
n(\nu)=\left( B - \frac{\kappa \nu^2}{2(\kappa+2)}\right)^{1/\kappa}\;.
\end{equation}
Here $B$ is an integration constant (see also \cite{nonlineardiffusion}).

For the case of strong chaos $\kappa=2$ and $\alpha=1/2$ in accord with 
the above results. Also for $\kappa=4$ we obtain $\alpha=1/3$ again in agreement with the above results.

\section{Generalizations}

Let us consider $\boldsymbol{D}$-dimensional lattices with
nonlinearity order $\sigma > 0$:
\begin{equation}
i\dot{\psi_{\boldsymbol{l}}}= \epsilon_{\boldsymbol{l}} \psi_{\boldsymbol{l}}
-\beta |\psi_{\boldsymbol{l}}|^{\sigma}\psi_{\boldsymbol{l}}
-\sum\limits_{\boldsymbol{m}\in
D(\boldsymbol{l})}\psi_{\boldsymbol{m}}\;.
\label{RDNLS-EOMG}
\end{equation}
Here $\boldsymbol{l}$ denotes an $\boldsymbol{D}$-dimensional lattice vector with
integer components, and $\boldsymbol{m}\in
D(\boldsymbol{l})$ defines its set of nearest neighbour lattice sites.
We assume that (a) all NMs are spatially localized (which can be obtained for strong
enough disorder $W$), (b) the property $\mathcal{W}(x \rightarrow 0) \rightarrow const \neq 0$
holds, and (c) the probability of resonances on the edge surface of a wave packet is tending
to zero during the spreading process. A wavepacket with average norm $n$ per excited mode has a second moment
$m_2 \sim 1/n^{2/\boldsymbol{D}}$. The nonlinear frequency shift is proportional to $\beta n^{\sigma/2}$.
The typical localization volume of a NM is still denoted by $V$, and the average spacing by $d$.

Consider a wave packet with norm density $n$ and volume $L < V$. A straightforward generalization of
the expected regimes of spreading leads to the following:
\begin{eqnarray}
\beta n^{\sigma/2} \left( \frac{L}{V}\right)^{\sigma/2} V   < \Delta \;:\; {\rm weak\; chaos}
\nonumber
\\
\beta n^{\sigma/2} \left( \frac{L}{V}\right)^{\sigma/2} V > \Delta \;:\; {\rm strong\; chaos}
\nonumber
\\
\beta n^{\sigma/2} > \Delta \;:\; {\rm selftrapping}
\nonumber
\end{eqnarray}
The regime of strong chaos, which is located between selftrapping and weak chaos,
can be observed only if 
\begin{equation}
L > L_c = V^{1-2/\sigma}\;,\; n > n_c = \frac{V}{L} \left( \frac{d}{\beta}\right)^{2/\sigma}\;.
\end{equation}
For $\sigma =2$ we need $L>1$, for $\sigma \rightarrow \infty$ we need $L > V$,
and for $\sigma < 2$ we need $L \geq 1$. Thus the regime of strong chaos can be observed
e.g. in a one-dimensional system with a single site excitation and $\sigma < 2$.

If the wave packet size $L > V$ then the conditions for observing different regimes simplify to
\begin{eqnarray}
\beta n^{\sigma/2} < d \;:\; {\rm weak\; chaos}
\nonumber
\\
\beta n^{\sigma/2} > d \;:\; {\rm strong\; chaos}
\nonumber
\\
\beta n^{\sigma/2} > \Delta \;:\; {\rm selftrapping}
\nonumber
\end{eqnarray}
The regime of strong chaos can be observed if 
\begin{equation}
n > n_c = \left( \frac{d}{\beta}\right)^{2/\sigma}\;.
\end{equation}

Similar to the above we obtain a diffusion
coefficient
\begin{equation}
D \sim \beta^2 n^{\sigma} (\mathcal{P}(\beta n^{\sigma/2}))^2
\;.
\label{ggeneralizeddiffusion}
\end{equation}
In both regimes of strong and weak chaos the spreading is subdiffusive
\cite{fks08}:
\begin{eqnarray}
m_2 \sim (\beta^2 t)^{\frac{2}{2+\sigma \boldsymbol{D}}}\;,\;{\rm strong}\;{\rm chaos}\;,
\label{sigma_strong}
\\
m_2 \sim (\beta^4 t)^{\frac{1}{1+\sigma \boldsymbol{D}}}\;,\;{\rm weak}\;{\rm chaos}\;.
\label{sigma_weak}
\end{eqnarray}

Let us calculate the number of resonances in the wave packet volume ($N_{RV}$) and on its surface ($N_{RS}$)
in the regime of weak chaos:
\begin{equation}
N_{RV} \sim \beta n^{\sigma/2-1}\;,\; N_{RS} \sim \beta n^{\frac{\boldsymbol{D}(\sigma-2)+2}{2\boldsymbol{D}}}\;.
\end{equation}
We find that there is a critical value of nonlinearity power $\sigma_c = 2$ such that
the number of volume resonances grows for $\sigma < \sigma_c$ with time, drops for $\sigma > \sigma_c$ and
stays constant for $\sigma=\sigma_c$. Therefore subdiffusive spreading will be more effective for $\sigma < \sigma_c$.

We also find that the number of surface resonances will grow with time for
\begin{equation}
\boldsymbol{D} > \boldsymbol{D_c}=\frac{1}{1-\sigma/2}\;,\; \sigma < 2\;.
\end{equation}
Therefore, for these cases, the wave packet surface will not stay compact. Instead surface resonances will lead to
a resonant leakage of excitations into the exterior. This process will increase the surface area, and therefore
lead to even more surface resonances, which again increase the surface area, and so on. 
The wave packet will fragmentize, perhaps get a fractal-like structure, and
lower its compactness index. The spreading of the wave packet will speed up, but will not anymore be due to
pure incoherent transfer, instead it will become a complicated mixture of incoherent and coherent transfer processes.

Mulansky computed spreading exponents for single site excitations with $\beta=1$, $W=4$,
$L=1$, $\boldsymbol{D}=1$ $n=1$ and $\sigma =1,2,4,6$ \cite{mm09}. Since for
$\sigma=2,4,6$ strong chaos is avoided, the fitting of the dependence $m_s(t)$ with a single power law is reasonable.
The corresponding fitted exponents $0.31\pm0.04$ ($\sigma=2$), $0.18 \pm 0.04$ ($\sigma=4$) and 
$0.14 \pm 0.05$ ($\sigma=6$) agree well with the predicted weak chaos result $1/3,1/5,1/7$ from (\ref{sigma_weak}).
For $\sigma=1$ the initial condition is launched in the regime of strong chaos. A single power law fit will therefore
not be reasonable. Since the outcome is a mixture of first strong and later possibly weak chaos,
the fitted exponent should be a number which is located between the two theoretical values $1/2$ and $2/3$.
Indeed, Mulansky reports a number $0.56\pm 0.04$ confirming our prediction.
Veksler et al \cite{hvyksf09} considered short time evolutions of single site excitations (up to $t=10^3$).
While the time window may happen to be too short for conclusive results, the observed increase
of fitted exponents with increasing $\beta$ for $\sigma < 2$ is possibly also influenced by the crossover
from weak to strong chaos.
Note that Skokos et al \cite{sf10} performed a more detailed analysis for the KG lattice, which 
confirm many the above results. 

\section{Discussion and conclusions}

Let us summarize the findings. If the strength of nonlinearity is large enough, a wave packet (or at least
an appreciable part of it) is selftrapped due to the finite bounds for the spectrum of the linear equation.
If the nonlinearity is weak enough so as to avoid selftrapping, two possible outcomes are predicted, which now
depend also on the volume $L$ of the packet. If $L > L_c$ and $n > n_c$, the NMs in the packet will be all resonant,
strongly interacting with each other and quickly dephase. That leads to a regime of strong chaos.
As time grows, the norm density $n$ drops below $n_c$, and the spreading continues in the regime
of weak chaos. If however either $L < L_c$ or $n < n_c$, strong chaos is avoided, and the packet
will spread in the regime of weak chaos. Lowering $\beta$ or $n$ further will keep the spreading
in the regime of weak chaos, but time scales of subdiffusion will grow, and the process will 
not be observable on the finite time window currently accessible by computational experiments.
The above holds if $\boldsymbol{D} < \boldsymbol{D_c}$ which implies that Anderson localization is preserved
in the tails and destroyed in the wave packet core. In other words, the time scales for destroying
Anderson localization in the tails are much larger than the time scales which lead to a thermalization
of the core and the corresponding subdiffusive spreading.
In order to observe the crossover from strong to weak chaos, one has to carefully choose the
system parameters. In particular, it is desirable to make the crossover region more narrow.

If $\boldsymbol{D} > \boldsymbol{D_c}$ then the spreading process will be different from the above predictions,
because resonant interaction in the surface and the tails of the wave packet will destroy Anderson
localization as well. The spreading will presumably stay subdiffusive. But we do not know currently
how to estimate and characterize the details of this process.

Our results rely on a conjecture of the dependence of a diffusion coefficient on the probability of
resonances. Future investigations may consider the connection between this conjecture and
the dependence of Lyapunov coefficients, relaxation times of correlation functions, and detrapping times
on the system parameters.
\\
Acknowledgements
\\
I thank I. Aleiner, B. Altshuler, S. Aubry, J. Bodyfelt, S. Fishman, D. Krimer, Y. Krivolapov, T. Lapteva, N. Li, Ch. Skokos, 
and H. Veksler for useful discussions.

%% The Appendices part is started with the command \appendix;
%% appendix sections are then done as normal sections
%% \appendix

%% \section{}
%% \label{}

%% References
%%
%% Following citation commands can be used in the body text:
%% Usage of \cite is as follows:
%%   \cite{key}         ==>>  [#]
%%   \cite[chap. 2]{key} ==>> [#, chap. 2]
%% 

%% References with bibTeX database:

\bibliographystyle{elsarticle-num}
%\bibliography{<your-bib-database>}

%% Authors are advised to submit their bibtex database files. They are
%% requested to list a bibtex style file in the manuscript if they do
%% not want to use elsarticle-num.bst.

%% References without bibTeX database:

% \begin{thebibliography}{00}

%% \bibitem must have the following form:
%%   \bibitem{key}...
%%

% \bibitem{}

% \end{thebibliography}

\end{document}